
\magnification\magstephalf
\tolerance 10000

\font\rfont=cmr10 at 10 true pt
\def\ref#1{$^{\hbox{\rfont {[#1]}}}$}


\font\fourteenbf=cmbx12 scaled\magstep1

\font\twelverm=cmr12

   
\def\e{\epsilon}  \def\o{\omega}
  \def\la{\lambda}
\def\T {\Theta}  
\def\pd {\partial}
\def\pmb#1{\setbox0=\hbox{#1}
 \kern.05em\copy0\kern-\wd0 \kern-.025em\raise.0433em\box0 }


 %


\def\boxit#1{\vbox{\hrule\hbox{\vrule\kern1pt\vbox
{\kern1pt#1\kern1pt}\kern1pt\vrule}\hrule}}

\parskip=6pt
\parindent=0pt
\hsize=17truecm\hoffset=-5truemm
\voffset=-1truecm\vsize=26truecm
\def\footnoterule{\kern-3pt
\hrule width 17truecm \kern 2.6pt}


\catcode`\@=11 

\def\nolabels{\def\wrlabeL##1{}\def\eqlabeL##1{}\def\reflabeL##1{}}
\def\writelabels{\def\wrlabeL##1{\leavevmode\vadjust{\rlap{\smash%
{\line{{\escapechar=` \hfill\rlap{\sevenrm\hskip.03in\string##1}}}}}}}%
\def\eqlabeL##1{{\escapechar-1\rlap{\sevenrm\hskip.05in\string##1}}}%
\def\reflabeL##1{\noexpand\llap{\noexpand\sevenrm\string\string\string##1}}}
\nolabels
\global\newcount\refno \global\refno=1
\newwrite\rfile
\def\defref{$^{{\hbox{\rfont [\the\refno]}}}$\nref}
\def\nref#1{\xdef#1{\the\refno}\writedef{#1\leftbracket#1}%
\ifnum\refno=1\immediate\openout\rfile=refs.tmp\fi
\global\advance\refno by1\chardef\wfile=\rfile\immediate
\write\rfile{\noexpand\item{#1\ }\reflabeL{#1\hskip.31in}\pctsign}\findarg}
\def\findarg#1#{\begingroup\obeylines\newlinechar=`\^^M\pass@rg}
{\obeylines\gdef\pass@rg#1{\writ@line\relax #1^^M\hbox{}^^M}%
\gdef\writ@line#1^^M{\expandafter\toks0\expandafter{\striprel@x #1}%
\edef\next{\the\toks0}\ifx\next\em@rk\let\next=\endgroup\else\ifx\next\empty%
\else\immediate\write\wfile{\the\toks0}\fi\let\next=\writ@line\fi\next\relax}}
\def\striprel@x#1{} \def\em@rk{\hbox{}}
\def\lref{\begingroup\obeylines\lr@f}
\def\lr@f#1#2{\gdef#1{\defref#1{#2}}\endgroup\unskip}
\newwrite\lfile
{\escapechar-1\xdef\pctsign{\string\%}\xdef\leftbracket{\string\{}
\xdef\rightbracket{\string\}}}

\def\writestop{\def\writestoppt{\immediate\write\lfile{\string\p
ageno%
\the\pageno\string\startrefs\leftbracket\the\refno\rightbracket%
\string\def\string\secsym\leftbracket\secsym\rightbracket%
\string\secno\the\secno\string\meqno\the\meqno}\immediate\closeout\lfile}}
\def\writestoppt{}\def\writedef#1{}
\catcode`\@=12 

\def\T{{\cal T}}
{\nopagenumbers
\line{{\fourteenbf }\hfill DAMTP 94/2}
\vskip 1truecm
\centerline{\fourteenbf PHOTON RADIATION IN A HEAT BATH}
\bigskip
\centerline{\twelverm P V Landshoff and J C Taylor}
\vskip 6pt
\centerline{DAMTP, University of Cambridge}
\bigskip
{\bf Abstract}

We discuss the bremsstrahlung of photons into a heat bath, and calculate
from first principles the energy radiated. Even to lowest order the
spectrum of the radiation at low frequency is no more singular than at
zero temperature. In addition to the obvious contributions, this spectrum
includes terms associated with fluctuations.
\bigskip
\bigskip
In a recent paper, Weldon\defref\weldon{
H A Weldon, West Virginia preprint WVU-932
} has considered the spectrum of photons emitted
when a charged particle scatters off a potential within a heat bath.
As he observes, in such a problem infrared effects are more delicate than
at zero temperature, because in addition to the usual factors the integrands
involved contain Bose distributions
$$
f(k^0)={1\over{e^{|k^0|/T}-1}}
\eqno(1)
$$
which diverge at zero energy. According to Weldon,
while the infrared divergences
do nevertheless still cancel in a lowest-order perturbative calculation
(this is an extension of the zero-temperature Bloch-Nordsieck mechanism),
they leave behind a spectrum of emitted photons whose number density
$dN/d\omega$ at zero-frequency
$\o$ is unacceptable: it behaves as $\o ^{-2}$,
so that the energy density $\omega dN/d\omega$
behaves as $\o ^{-1}$, which integrates
to give a divergent result for the total energy carried off by the
photons\footnote{$^*$}{See equation (1.7) of reference \weldon}.

Weldon further shows how to sum the infrared-sensitive terms to all
orders.
In this note, we are not concerned with this, the most important part
of Weldon's beautiful paper. Rather, we seek to clarify his remarks about the
lowest-order calculation of the radiated energy.
We do so within the framework
of a very simple model\defref\jacob{
M Jacob and P V Landshoff, Physics Letters B281 (1992) 114
},
which we believe incorporates the essential
features. This is the decay, within a heat bath that contains only photons,
of a scalar uncharged particle $P$ into a pair of scalar charged particles
(``muons'').
We calculate three different quantities: (I) the order-$e^2$ contribution
$\Delta\Gamma$ to the total decay rate; (II) the probability that the muons
emerge with combined centre-of-mass
energy within $\Delta E$ of the parent-particle
energy; and (III) the net energy spectrum $\omega dN/d\omega$
transferred to the heat bath by the photons.

For (I), we agree with Weldon that $\Delta\Gamma$ is finite. For (II), we
find that the probability in order $e^2$ behaves as $(\Delta E)^{-1}$;
we expect that Weldon's resummation procedure would remove this singular
behaviour. It is for (III) that we differ from Weldon: $\omega dN/d\omega$
is finite at $\omega =0$ even to order $e^2$.

As is well-known, the decay rate may be calculated from a discontinuity of
a thermal Green's function\defref\disc{
H A Weldon, Physical Review D42 (1990) 2384;
R Kobes and G W Semenoff, Nuclear Physics B272 (1986) 329;
N Ashida et al, Physical Review D45 (1992) 2066
}.
However, it appears that the energy spectrum $\omega dN/d\omega$ cannot be
calulated in terms of such a discontinuity. Instead, we must introduce a
different method,
which we derive from first principles. This method reproduces the same
result for the decay rate as the conventional one, and may be used also
for calculating $\omega dN/d\omega$.

We recall first that in finite-temperature gauge theory one may choose
that at bare-propagator level
only the physical degrees of freedom are thermalised\defref\rebhan{
P V Landshoff and A Rebhan, Nuclear Physics B383 (1992) 607
}.
Then the thermal part of the photon propagator
$$
\Delta _T (k)=2\pi\delta (k^2)f(k)
\eqno(2a)
$$
\vfill\eject
}\pageno=2
represents the effects of real photons in the heat bath. This is easy
to understand.
The finite-temperature photon propagator is, in any gauge,
$$
Z^{-1}\hbox{tr }e^{-H_0/T} TA(x)A(0)
=Z^{-1}\hbox{tr }e^{-H_0/T}\big (\langle 0|TA(x)A(0)|0\rangle
          +:A(x)A(0):\big )
\eqno(3a)
$$
where
$$
Z=\hbox{tr }e^{-H_0/T}
\eqno(3b)
$$
and the traces are defined as
summation over expectation values in only {\it physical}
states.
The first term in (3a) is just the zero-temperature Feynman propagator
in whatever gauge is being used,
and is unaltered by
the heat bath. The second receives contributions only from the
real photons in the heat bath and is
$$
Z^{-1}\hbox{tr }e^{-H_0/T}
\int {{d^4k}\over{(2\pi)^4}}2\pi\delta ^{+}(k^2)
{{d^4k'}\over{(2\pi)^4}}2\pi\delta ^{+}(k'^2)
\left \{ a^{\dag}(k)a(k')e^{ik.x}+a^{\dag}(k')a(k)e^{-ik.x}\right \}
\eqno(2b)
$$
which is the  Fourier transform of (2a).

We return now to the decay in the heat bath of the uncharged particle
into two muons.
We have previously\ref{\jacob} calculated,
to lowest order $e^2$
in the muon charge, the change $\Delta\Gamma$
in the decay rate caused by the heat bath.
This change involves only the thermal part (2) of the photon propagator.
It consists of three pieces: (i) a term that may be interpreted as the
acquisition by the muons of an additional mass of order $eT$, (ii) another
that may be interpreted as a shift induced by the heat bath
in the coupling $\la$ that
governs the decay, and (iii) a term that corresponds to stimulated emission of
photons into the heat bath or absorption from it. We found that
the mass shift (i) is
$$
\delta m^2=2e^2\int {{d^4k}\over{(2\pi )^4}} 2\pi\delta (k^2)f(k^0)
\eqno(4a)
$$
and
the terms (ii) and (iii)
contribute
$$
2P^0\;\Delta \Gamma  = \lambda ^2e^2{1\over{(2\pi )^5}}
\int d^4k\; \delta (k^2)f(k^0)d^4p_1d^4p_2\;\delta ^{+}(p_1^2-m^2)
\delta ^{+}(p_2^2-m^2)
\left [- \left ({{p_1}\over{p_1.k}}-{{p_2}\over{p_2.k}}\right )^2\right ]
$$$$
\left\{ -\delta^4(p_1+p_2-P)+\delta^4(p_1+p_2+k-P)\right\}
\eqno(4b)$$
where in the last factor the first $\delta$-function corresponds to the
term (ii) and the second to (iii). We have supposed that the decaying
particle has momentum $P$.

In the integrals (2), $k$ denotes photon 4-momentum. Because
only the thermal part (2) of the photon propagator is involved.
the photon is real, even when its effect is
interpreted as just changing the value of the coupling $\la$, or
of the muon mass. These two changes together correspond to the
absorption of a real photon from the heat bath and its re-emission with
the same momentum.
The amplitude for this is the sum of
the order-$e^2$ graphs shown in figure 1a and 1b,
where the dashed lines at the bottom
are the spectator photons in the heat bath. Both of the active photons
in the graphs must have the same momentum $k$, because to obtain an order-$e^2$
contribution to the decay rate we must multiply the sum of graphs in figures
1a and 1b
by the order-$e^0$ graph of figure 1c, in which all the heat-bath photons are
spectators. This product corresponds to the first two
of the  of cut graphs shown in
figure 2, in which  the photon line represents the thermal
part (2a) of the photon propagator.
Because this thermal propagator contains $\delta (k^2)$, rather than
$\delta ^{+}(k^2)$, each single cut graph of figure 2 corresponds
to two contributions, as indicated in  in figures 1a and 1b.
The first $\delta$-function in the curly bracket in (4b) arises from
the first two cut graphs in figure 2, while the second $\delta$-function
arise from the other two cut graphs. The latter represent the sum
of the squares of the terms shown in figure 1d and 1e.

In his calculation of the net photon energy spectrum $\omega dN/d\omega$,
Weldon appears not to have
included the contributions of figure
1a, 1b and 1e. We maintain that it
is necessary to do so if one wishes to calculate
a physically-meaningful quantity.  In order to construct such a quantity,
we must carefully perform a thought experiment. We shall consider two
such experiments. In both, we think of the heat bath as a microwave oven
with only photons in it. Any muons that are produced as a result of decays
escape straight
through the oven walls. Our world has no other charged particles, in particular
no electrons.

In our first thought experiment,
we drill a small hole in one of the oven walls and imagine
observing the energy spectrum $\omega dN/d\omega$ of
radiation emitted through it, just as one does in the usual
analysis of black-body radiation\defref\mandl{
eg F Mandl, {\sl Statistical Physics}, Wiley 1971
}.  We suppose that we have  introduced into the oven a
collection of the particles $P$ that decay into two muons,
and consider the  spectrum of photons in events triggered by the
decay-product muons. We
calculate the difference between this photon spectrum at
temperature $T$ and that at zero temperature.

In order to do this calculation, we go back to first principles.
Define single-photon states $r$, initially discrete
with energies $\e _r$, and label the
states of the heat bath
by the corresponding occupation numbers $n_r$. Then, with a
decaying particle $P$ also in the oven, the initial density matrix
is
$$\rho _i=Z^{-1} \sum _{\{ n\} }\exp\big ({-\sum _rn_r\e _r/T}\big )
\;|n_1,n_2,\dots ,P\rangle
\langle n_1,n_2,\dots ,P|
\eqno(5a)
$$
with
$$
Z=\prod _r{1\over{1-e^{-\e _r/T}}}
\eqno(5b)
$$
After a long time, the density matrix has become
$$\rho _f=Z^{-1} \sum _{ \{n\}}\exp\big ({-\sum _rn_r\e _r/T}\big )
\;\T|n_1,n_2,\dots ,P\rangle
\langle n_1,n_2,\dots ,P|\T^{\dag}
\eqno(6a)
$$
Here, as usual, $\T$ is the interaction part of the $S$-matrix, $S=1+i\T$.
Because our initial state contains an unstable particle, which inevitably
decays, we know that an interaction takes place.
The decay rate of the particle $P$ is calculated from the expectation value
of $\rho _f$ in the states
$$
| m_1,m_2,\dots \mu (p_1) \mu (p_2)\rangle
\eqno(6b)
$$
with summation over the photon occupation numbers $ m_1,m_2,\dots$ and
over the momenta $p_1$ and $p_2$ of the decay-product muons.
For this, we need the photon-operator structure of the $\T$-matrix.
The various matrix elements in figure 1 correspond to the following
photon-operator terms in $\T$:
$$
e^2\sum _s A_sa^{\dag}_sa_s+e^2\sum _s B_sa_sa^{\dag}_s
+C+e\sum _s D_sa^{\dag}_s+e\sum _s E_sa_s
\eqno(7)
$$
The operators $A_s,B_s, \dots$ correspond to figures 1a,b,\dots\ and
involve the photon modes $s$ only through
conserving energy and momentum. They convert the initial
particle $P$ into the two muons and include also the
internal muon propagators.
We use the familiar properties $a|n\rangle =\sqrt{n}\;|n-1\rangle$ and
$a^{\dag}|n\rangle =\sqrt{n+1}\;|n+1\rangle$, so that the
order-$e^2$ contribution to the summed expectation
value takes the form
$$
\Delta\Gamma =Z^{-1}\sum _{\{ n\} }\exp\big (-\sum _rn_r\e _r/T\big )\sum
_{s,p_1,p_2}
\big (n_sP_s^{AC}+(n_s+1)P_s^{BC}+
(n_s+1)P_s^{DD}+n_sP_s^{EE}\big )
\eqno(8a)
$$
Here
$$
P_s^{AC}=e^2\langle \mu (p_1)\mu (p_2)|\;A_s\;|P\rangle
        \langle P|\;C^{\dag}\;|\mu (p_1)\mu (p_2)\rangle+
e^2\langle \mu (p_1)\mu (p_2)|\;C\;|P\rangle
        \langle P|\;A_s^{\dag}\;|\mu (p_1)\mu (p_2)\rangle
\eqno(8b)
$$
and similarly for $P_s^{BC},P_s^{DD}$ and $P_s^{EE}$.
Notice that, in this order-$e^2$ expression for $\Delta\Gamma$, the $DD$
and $EE$ terms are the squares of order-$e$ amplitudes, while the $AC$ and
$BC$ terms are the interference between order-$e^2$ and order-$e^0$
amplitudes. The presence of the interference terms at finite temperature
is possible because the order-$e^0$ amplitude in figure 1c involves
spectator photons in the heat bath. One of these spectator photons has to be
identified with the the photon that is absorbed and emitted in the diagrams
of figure 1a and 1b. For this to be possible, the absorbed and emitted photon
must have the same momentum as the corresponding spectator photon in figure
1c, and so the photon scattering in figures 1a and 1b is forward scattering.

We  use
$$
\bar n \equiv Z^{-1}\sum _n n\e^{-n\e /T}=f(\e )
\eqno(9)
$$
where $f(\e )$ is the Bose distribution (1), so that (8a) is
$$
\Delta\Gamma =\sum _{s,p_1,p_2}
\big (f(\e _s)P_s^{AC}+(f(\e _s)+1)P_s^{BC}+
(f(\e _s)+1)P_s^{DD}+f(\e _s)P_s^{EE}\big )
\eqno(10)
$$
Subtracting off the zero-temperature limit, we retrieve in the continuum limit
the formula (4b).

We now calculate the thermal average of the number of photons in state $t$.
This comes from
the expectation value of $a_t^{\dag}a_t\rho _f$ in the states (6b),
again with summation over the photon occupation numbers $ m_1,m_2,\dots$ and
over the momenta $p_1$ and $p_2$ of the decay-product muons. Because $\rho _f$
is not normalised to unity, we must introduce also a factor
$\Gamma ^{-1}$:
$$
\langle N_t \rangle =\Gamma ^{-1}
Z^{-1}\sum _{\{ n\} }\exp\big (-\sum _rn_r\e _r/T\big )\sum _{p_1,p_2}
\Big \{ n_t^2P_t^{AC}+n_t(n_t+1)P_t^{BC}+
(n_t+1)^2P_t^{DD}+(n_t-1)n_tP_t^{EE}
$$$$
+n_tP^{CC}+n_t
\sum _{s\not= t}\big (n_sP_s^{AC}+(n_s+1)P_s^{BC}+
(n_s+1)P_s^{DD}+n_sP_s^{EE}\big )
\Big \}
\eqno(11)
$$
We use
$$
\overline{n^2}\equiv Z^{-1}\sum _n n^2\e^{-n\e /T}=\bar n^2
+f(\e )\big ( f(\e )+1\big )
\eqno(12)
$$
so that (11) is
$$
\langle N_t \rangle =f(\e _t)+F^{\hbox{{\sevenrm emission}}}(\e _t)
+F^{\hbox{{\sevenrm absorption}}}(\e _t)
+F^{\hbox{{\sevenrm fluc}}}(\e _t)
\eqno(13a)
$$
where
$$
F^{\hbox{{\sevenrm emission}}}(\e _t)=
\Gamma ^{-1}\sum _{p_1,p_2}\big (f(\e _t)+1\big  )P_t^{DD}
$$
$$
F^{\hbox{{\sevenrm absorption}}}(\e _t)=
-\Gamma ^{-1}\sum _{p_1,p_2}f(\e _t)P_t^{EE}
$$
$$
F^{\hbox{{\sevenrm fluc}}}(\e _t)=\Gamma ^{-1}\sum _{p_1,p_2}
 f(\e _t)(f(\e _t)+1)\big [ P_t^{AC}+P_t^{BC}+P_t^{DD}+P_t^{EE}\big ]
\eqno(13b)
$$

Evidently, the first term in (13a) is the  value $\langle N_t \rangle$
would have if there was no decaying particle in the heat bath. We subtract
it off. As expected, the emission term, corresponding to part of the square
of figure 1d, adds to the spectrum, while the absorption term,
corresponding to part of the square of figure 1e, has negative sign and
so reduces it.
{}From (12), the term $F^{\hbox{{\sevenrm fluc}}}(\e _t)$, because it
involves $f(f+1)$,
is associated with fluctuations in the heat bath.

Subtracting off
the zero-temperature limit, and going over to the continuum limit, we
are left with the change in the spectrum per decaying particle:
$$
2P^0\;\Gamma {{d}\over{d\o}}\Delta N(\o )=\Delta ^{\hbox{{\sevenrm emission}}}
+\Delta ^{\hbox{{\sevenrm absorption}}}+
\Delta ^{\hbox{{\sevenrm fluc}}}
\eqno(14a)
$$
with
$$
\Delta ^{\hbox{{\sevenrm emission}}}+\Delta ^{\hbox{{\sevenrm absorption}}}
=f(\o )
{{\lambda ^2e^2}\over{(2\pi )^5}}
\int d^4k \delta (k^0-\o )\delta (k^2)d^4p_1d^4p_2
\delta ^{+}(p_1^2-m^2)\delta ^{+}(p_2^2-m^2) ~~~~~
$$$$
{}~~~~~~~~~~~~~~~~~J^2 \Big\{ \delta^4(p_1+p_2+k-P)
-\delta^4(p_1+p_2-k-P)\Big\}
\eqno(14b)
$$
and
$$
\Delta ^{\hbox{{\sevenrm fluc}}}=
f(\o )(f(\o )+1)
{{\lambda ^2e^2}\over{(2\pi )^5}}
\int d^4k \delta (k^0-\o )\delta (k^2)d^4p_1d^4p_2
\delta ^{+}(p_1^2-m^2)\delta ^{+}(p_2^2-m^2) ~~~~~~~~~
$$$$
{}~~~~~~~~~~~~~~~~~J^2
 \Big\{ \delta^4(p_1+p_2+k-P)
+\delta^4(p_1+p_2-k-P)
-2\delta^4(p_1+p_2-P)\Big\}
$$$$
+f(\o )(f(\o )+1){{4\lambda ^2e^2}\over{(2\pi )^5}}
\int d^4k \delta (k^0-\o )\delta (k^2)
{\pd\over \pd m^2}\int d^4p_1d^4p_2
\delta ^{+}(p_1^2-m^2)\delta ^{+}(p_2^2-m^2)
 \delta^4(p_1+p_2-P)
\eqno(14c)
$$
where
$$
J^2=\left [-\left ({{p_1}\over{p_1.k}}-{{p_2}\over{p_2.k}}\right )^2\right ]
\eqno(14d)
$$
The last term in (14c) is the  part of the interference between figures 1a and
1b with figure 1c that  is associated with the thermal mass shift $\delta m^2$
of (4a).

When $\o\to 0$, ${{d}\over{d\o}}\Delta N(\o )$
diverges as $\o ^{-1}$. The $\o ^{-2}$
mentioned  by Weldon\ref{\weldon} is present in
$F^{\hbox{{\sevenrm absorption}}}$ and $F^{\hbox{{\sevenrm emission}}}$,
but it cancels between them and leaves behind $\o ^{-1}$, The
separate contributions to the fluctuation term behave as $\o ^{-3}$.
An important component of the fluctuation term is the interference between
the diagrams of figures 1a and 1b with figure 1c; the $\delta m^2$
term is a part of this interference term\ref{\jacob}.
It is straightforward to see that the $\o ^{-3}$ behaviour cancels
between the various parts of the fluctuation term, as does its
$\o ^{-2}$ behaviour.
Moreover, one can show that, no matter how fast the decaying particle is
moving (or not moving) before it decays,
the $\o ^{-1}$ behaviour cancels also; this cancellation is closely
related to the cancellation that occurs at low temperature in
$\Delta \Gamma$, making\ref{\jacob} it behave as $T^4$ instead of $T^2$.
As we show in the Appendix, it is a result of current conservation
together with unitarity. So the fluctuation term is nonsingular when
$\o\to 0$; it actually goes to zero like $\o$. However, notice that
the $\o ^{-1}$ behaviour in the fluctuation term cancels only because we
have integrated over the momenta $p_1$ and $p_2$
of the final-state charged particles. In the scattering problem considered by
Weldon\ref{\weldon}, $p_1$ rather is the momentum of an incoming particle
and is not integrated,
so we believe that then the cancellation of the $\o ^{-1}$
contributions to the fluctuation terms does not take place.

Thus the breakdown of perturbation theory signalled by Weldon in fact
does not occur for the measurement he considered. However, it does
occur in certain other cases. To see this,
consider our second thought experiment (II), where instead of measuring the
number density of the photons emitted from the heat bath we measure the
total energy of the two muons. We consider the rate at which the pair of muons
emerges with total energy within magnitude $\Delta E$ of the energy of their
parent particle. This is calculated from (4b) with an additional
$$
\theta \big{(}\Delta E -|P^0-p_1^0-p_2^0|\big{)}
\eqno(15)
$$
under the integral. In spite of the presence in (4b) of the factor that
is quadratically divergent at $k=0$, the $k$-integration is convergent,
because although the difference of $\delta$-functions
in the last factor is linear
in $k$, the rest of the integrand is even in $k$. However, although
this cancellation of divergences takes place between the two
$\delta$-functions,
as $\Delta E \to 0$ the answer is again divergent.
The term with the first $\delta$-function clearly is
independent of $\Delta E $, while for small $\Delta E$ the second behaves as
$\Delta E ^{-1}$.
This behaviour is unphysical (it is even large negative)
and so must be corrected by a resummation such as was suggested by
Weldon\ref{\weldon}. Of course, the same resummation is necessary for
this quantity at zero temperature, though then before the resummation
the divergence is only as $\log\Delta E$.

We have explained that our method of calculation requires us to go back
to first principles. It gives the same answer for the decay rate as
the conventional approach, but may be applied also to the calculation
of the photon energy spectrum $\omega dN/d\omega$ or the number spectrum
$dN/d\omega$. While the decay rate is linear in the Bose distribution
$f(\omega )$, the number spectrum $dN/d\omega$ contains a quadratic term,
which is associated with fluctuations in the heat bath. Therefore, it
does not seem to be associated with a discontinuity of a thermal Green's
function. Our initial attempts to calculate $dN/d\omega$
from the discontinuity diagrams of figure 2 yielded an absorption term
whose sign was opposite from that in (13b), which is clearly incorrect;
as we have shown, the correct sign is obtained only when one takes account
of the fact that part of the absorption effect is a contribution to the
fluctuation term.

\bigskip
{\sl We are grateful to Arthur Weldon for a very useful correspondence.
This research is supported in part by the EC Programme "Human Capital
and Mobility", Network "Physics at High Energy Colliders", contract
CHRX-CT93-0537 (DG 12 COMA)
}

\bigskip
\bigskip
{\bf Appendix}

In this Appendix, we calculate the leading terms in the photon spectrum
(14) for small $\omega$.
We verify that the $\o ^{-2}$ terms cancel in the emission and
absorption parts of (14b), leaving $\o ^{-1}$.
We find further that there is a cancellation between the different parts of
the fluctuation term (14c), so that while the separate terms behave as
$\o ^{-3}$ their sum behaves as $\o$.
This is closely related to the cancellation, found in [\jacob ],
of the low-temperature contributions to $\Delta \Gamma$. As we explain at the
end of this Appendix, it is a consequence of current conservation and
unitarity.

In order to deal with (14), we follow reference~{\jacob}~and
make the changes of variables
$$
p_1 \rightarrow p_1 \mp {p_2.k \over P.k}k,~~~p_2 \rightarrow p_2 \mp
{p_1.k \over P.k}k \eqno(A1)
$$
so that
$$
J^2\;\delta ^{+}(p_1^2-m^2)\delta ^{+}(p_2^2-m^2)
\delta^4(p_1+p_2-P \pm k)\longrightarrow
J^2\;\delta^{+}(p_1^2-m^2 \mp X)\delta^{+}(p_2^2 -m^2 \mp X)
\delta^4(p_1+p_2-P)
\eqno(A2)
$$
where
$$
X = {2p_1.k\; p_2.k \over P.k}
\eqno(A3)
$$
We then expand (A2) in powers of $X$, to the second order for the
fluctuation  contribution to (14), and to first order for the
emission and absorption contributions.
The result is
$$
\Delta ^{\hbox{{\sevenrm emission}}}+\Delta ^{\hbox{{\sevenrm absorption}}}
\sim
f(\o ){\lambda^2 e^2 \over (2\pi )^5}
\int d^4 k\delta(k^2) \delta(k^0 -\omega)A
\eqno(A4a)
$$
and
$$
\Delta ^{\hbox{{\sevenrm fluc}}}\sim
f(\o )(f(\o )+1)
{\lambda^2 e^2 \over (2\pi )^5}
\int d^4 k\delta(k^2) \delta(k^0 -\omega)B
\eqno(A4b)
$$
where
$$
A=\int d^4p_1d^4p_2\delta^4(P-p_1-p_2)(-2XJ^2)(\partial /\partial m^2)
\delta^{+}(p_1^2-m^2)\delta^{+}(p_2^2-m^2)
\eqno(A5a)
$$
and
$$
B=\int d^4p_1 d^4p_2\delta^4 (P-p_1 -p_2)\;[X^2J^2(\partial /\partial m^2)^2
+4(\partial /\partial m^2)]\;\delta^{+}(p_1^2-m^2)\delta^{+}(p_2^2-m^2),
\eqno(A5b)
$$
with $J^2$ defined in (14d).

Evidently both $A$ and $B$ are Lorentz invariants and so
they can depend upon $k$
via the invariant $P.k$ only. Therefore we can first calculate with
the $P$-particle at rest (relative to the heat bath), and then replace
$M\omega$ by
$P.k$ to get the result for a moving $P$-particle. For $P$ at rest,
and with $\theta$ the angle between {\bf k} and {\bf p}$_1$,
$$
{1 \over 2}\int^1_{-1} d(\cos\theta)\;XJ^2 ={2M \over \omega}
\left [1-{2m^2 \over M\surd(M^2-4m^2)}
\ln \left ({M+\surd(M^2-4m^2) \over M-\surd(M^2-4m^2)} \right ) \right ]
\eqno(A6a)
$$
$$
{1 \over 2}\int^1_{-1} dx X^2J^2 ={2 \over 3}(M^2-4m^2)
\eqno(A6b)
$$
We have anticipated that we are going to insert these integrals into (A5) and
(A4), and so set $k^2=0$ and $p_1^2=m^2=p_2^2$.
We require also
$$
\int d^4p_1 d^4p_2\delta^4(P-p_1-p_2)\delta^{+}(p_1^2 -m^2)\delta^{+}
(p_2^2 -m^2)=\pi {\surd(M^2-4m^2)\over 2M}
\eqno(A6c)
$$

Since (A6b) and (A6c) do not depend on $\o$, they remain the same
when the decaying particle $P$ is not at rest in the heat bath.
Using (A6) we find that the two derivatives in the expression (A5b) for
$B$ give equal and opposite contributions when they are inserted into
(A4b). This cancellation is exactly similar to that which occurs
in the change $\Delta\Gamma$ induced by the heat bath
in the low-temperature decay rate\ref{\jacob}.

Thus $\Delta ^{\hbox{{\sevenrm fluc}}}$ vanishes
in this approximation: it receives its first contribution from
the $X^4$ terms in the expansion of $(A2)$ and so behaves as
$\o ^3f(\o )(f(\o )+1)$
at low frequency. The leading behaviour of
${{d}\over{d\o}}\Delta N(\o )$ near $\o =0$ comes from the
emission and absorption terms and is, from (A5a) and (A6a),
$$
2P^0\;\Gamma {d \Delta N \over d \omega} \sim {2\lambda^2 e^2 \over (2\pi)^3}
{f(\o )\over \surd({P^0}^2-M^2)}
\ln \left ({M+\surd(M^2-4m^2) \over M-\surd(M^2-4m^2)} \right )
\ln \left ({P^0+\surd({P^0}^2-M^2)\over P^0-\surd({P^0}^2-M^2)}\right )
\eqno(A7)
$$
where we have used
$$
\int d^4k \delta (k^0-\o )\delta (k^2) (P.k)^{-1}=
{2\pi\over \surd({P^0}^2-M^2)}
\ln \left ({P^0+\surd({P^0}^2-M^2)\over P^0-\surd({P^0}^2-M^2)}\right )
$$

The reason for the cancellation in $\Delta\Gamma$, and of the leading terms in
$B$, is the following.
$\Delta\Gamma$ is calculated from the trace of the final-state density
matrix $\rho _f$ given in (6a). Through unitarity,
this is
$$
\hbox{tr } Z^{-1} \sum _{ \{n\}}\exp\big ({-\sum _rn_r\e _r/T}\big )
\;\T|n_1,n_2,\dots ,P\rangle
\langle n_1,n_2,\dots ,P|\T^{\dag}~~~~~~~~~~~~~~
$$$$
{}~~~~~~~~~~~~~=
-iZ^{-1} \sum _{ \{n\}}\exp\big ({-\sum _rn_r\e _r/T}\big )
\langle n_1,n_2,\dots ,P|(\T-\T^{\dag})|n_1,n_2,\dots ,P\rangle
\eqno(A8)
$$
Because we are working to lowest order in the charge $e$, only one photon
in the state $|n_1,n_2,\dots ,P\rangle$ is active, with the other photons
just providing the number factors $n_s$ in (8a), so that in the
continuum limit $\Delta\Gamma$
is an integral, involving the Bose factor $f$,
over the photon momentum $k$ of the imaginary part
of the scattering amplitude $\langle P,k^{\mu}|\T|P,k^{\nu}\rangle$.
Because the electromagnetic current is conserved, for off-shell $k$
this amplitude has the usual decomposition:
$$
-\Big (g^{\mu\nu}-{k^{\mu}k^{\nu}\over k^2}\Big )T_1
+\Big ( P^{\mu}-{P.k\over k^2}k^{\mu}\Big )
\Big ( P^{\nu}-{P.k\over k^2}k^{\nu}\Big )T_2
\eqno(A9)
$$
This has to be contracted with transverse real-photon
polarisation vectors and $k^2$ set to 0. This eliminates $T_2$.
But when $k^2\to 0$, $T_1\sim (P.k)^2T_2/k^2$, and $T_2$ vanishes
like $k^2$. Because we do not allow the unstable particle $P$ to
survive in the final state,
the amplitudes $T_1$ and $T_2$ do not have poles corresponding to this
particle, and so $T_1$ behaves as $(P.k)^2$ for small $P.k$.
It is this
that makes $\Delta\Gamma$ behave\ref{\jacob} as $T^4$
at low temperature $T$. According to (8a), when the zero-temperature
limit has been subtracted off, $\Delta\Gamma$ is calculated from the
combination $\big [ P_t^{AC}+P_t^{BC}+P_t^{DD}+P_t^{EE}\big ]$. This same
combination appears in the fluctuation term (13b) and the $(P.k)^2$
there gives the additional factor $\o ^2$.

Note that it is essential to this argument that we have integrated
over the momenta $p_1$ and $p_2$ of the final-state muons.

\bigskip
\bigskip
\medskip\immediate\closeout\rfile\writestoppt
\baselineskip=14pt{{\bf References}}\bigskip{\frenchspacing%
\parindent=20pt\escapechar=` \input refs.tmp\bigskip}\nonfrenchspacing
\vfill\eject
\input pictex.tex
\voffset=8truemm
 \font\lab=cmr10 at 12 truept
\linethickness=1truemm
\setcoordinatesystem units <11truemm, 7truemm> point at 0 0
\beginpicture

\newbox\picA
\setbox\picA=\hbox{\beginpicture
\savelinesandcurves on "ird.a"
\setcoordinatesystem point at 0 0
\setsolid
\putrule from -2 0 to 0 0
\plot 0 0 2 2 /
\plot 0 0 2 -2 /
\setdashes
\plot -2 -2.6 2 -2.6 /
\plot -2 -3.0 2 -3.0 /
\endpicture}
\setdashes
\lab

\put{\copy\picA} at 9 0
\plot 8 -0.6 10.6 -0.6 /
\plot 11.2 1.2 12 1.2 /

\put{\copy\picA} at -1 0
\plot -2 -0.6 0.6 -0.6 /
\plot 1.2 -1.2 2 -1.2 /
\put{(a)} at 5 -4.5

\setcoordinatesystem point at 0 8
\put{\copy\picA} at -1 0
\plot -2 -1.2 1.2 -1.2 /
\plot 0.6 -0.6 2 -0.6 /

\put{\copy\picA} at 9 0
\plot 8 -1.2 11.2 -1.2 /
\plot 10.6 0.6 12 0.6 /
\put{(b)} at 5 -4.5

\setcoordinatesystem point at 0 16
\put{\copy\picA} at -1 0
\plot -2 -2,2 2 -2.2 /
\put{(c)} at 0 -4.5

\setcoordinatesystem point at 0 24
\put{\copy\picA} at -1 0
\plot -2 -2.2 2 -2.2 /
\plot 0.6 -0.6 2 -0.6 /

\put{\copy\picA} at 9 0
\plot 8 -1.2 11.2 -1.2 /
\put{(d)} at 0 -4.5
\put{(e)} at 10 -4.5

\endpicture
\vskip 1truecm
\font\ten=cmr10 scaled\magstephalf
\centerline{{\ten Figure 1}}
\bigskip
\parindent=0pt
{\ten
Photon radiation in a heat bath, involving both active and spectator photons.
}
\vfill\eject

\voffset=2truein
\setcoordinatesystem units <2truemm, 2truemm> point at 0 0
\beginpicture
\newbox\picA
\setbox\picA=\hbox{\beginpicture
\setcoordinatesystem point at 0 0
\setplotarea x from -17 to 17, y from -4 to 4
\linethickness=2pt
\putrule from -17 0 to -9 0
\putrule from 17 0 to 9 0
\ellipticalarc axes ratio 2:1 90 degrees from -9 0 center at -1 0
\ellipticalarc axes ratio 2:1 -90 degrees from -9 0 center at -1 0
\ellipticalarc axes ratio 2:1 90 degrees from 9 0 center at 1 0
\ellipticalarc axes ratio 2:1 -90 degrees from 9 0 center at 1 0
\put {$p_1$} at  -2 5.5
\put {$p_2$} at -2 -5.5
\put {$P$} at  -15 1.5
\put {$>$} at  -1 4
\put {$>$} at  -1 -4
\put {$>$} at  -13 0
\endpicture}
\setcoordinatesystem units <2truemm, 2truemm> point at 50 -20
\put{\copy\picA} at 0 0
\put {$<$} at -6.5 0.45
\put {$k$} at -6.1 1.9
\setdashes
\circulararc -110 degrees from -2.5 3.7 center at -6.5 4.5
\setsolid
\setcoordinatesystem units <2truemm, 2truemm> point at 10 -20
\put{\copy\picA} at 0 0
\put {$k$} at -4.5 0
\plot -6.5 .5 -6 -.2 -5.5 .5 /
\setdashes
\plot -6 3 -6 -3 /
\setsolid
\setcoordinatesystem units <2truemm, 2truemm> point at 50 0
\put{\copy\picA} at 0 0
\put {$k$} at -1.3 0
\put {$>$} at -1.3 1.75
\setdashes
\plot -6 3 -1 1.5 /
\plot 6 3 1 1.5 /
\setsolid
\setcoordinatesystem units <2truemm, 2truemm> point at 10 0
\put{\copy\picA} at 0 0
\put {$k$} at -2 -0.7
\plot -1.5 1.2 -1 0.5 -1.7 0.3 /
\setdashes
\plot -6 3 -1 .5 /
\plot 6 -3 1 -.5 /
\setsolid
\endpicture
\vskip 1truecm
\font\ten=cmr10 scaled\magstephalf
\centerline{{\ten Figure 2}}
\bigskip
\parindent=0pt
{\ten
Cut diagrams for the square of the amplitude corresponding to the
various terms of figure 1
}
\bye